\documentclass[aps,showpacs,twocolumn]{revtex4}
\usepackage{amssymb}
\usepackage{amsmath}
\usepackage{graphicx}
\usepackage{epsfig}

\begin{document}

\title{Perfect coherent shift of bound pairs in strongly correlated systems}
\author{L. Jin, Z. Song}
\email{songtc@nankai.edu.cn}
\affiliation{School of Physics, Nankai University, Tianjin 300071, China}

\begin{abstract}
In the present work we extend the concept of coherent shift for the extended
Bose-Hubbard model and Fermi-Hubbard model. We present two types of local
bound pair (BP) for Bose system and one type for Fermi system. It is shown
exactly that the perfect coherent shift can be achieved in such models. We
find that for a Bose on-site BP, the perfect coherent shift condition
depends on the nearest-neighbor interaction strength and the momentum of the
incident single particle wavepacket, while for the other two types of BPs,
it is independent of the initial state in the proposed systems.
\end{abstract}

\maketitle


\section{Introduction}

Bound pair (BP) state and its dynamics are interesting recent topics in
quantum physics and quantum information \cite{Winkler, Mahajan, Petrosyan,
Creffield, Kuklov, Folling, Zollner, ChenS, Valiente, JLBP, MVExtendB}. In a
previous work \cite{JLBP}, we studied the dynamics of a BP and the
interaction between a single particle and a BP. Within the large $U$ regime,
we first found an interesting scattering process, coherent shift, between
them. We believed that, this phenomenon should not be exclusive and could be
applied to quantum device design. In addition, M. Valiente \textit{et al}.
performed a comprehensive numerical simulation of such a scattering process,
which is useful to understand this phenomenon \cite{Comment}. It was pointed
that a perfect coherent shift is hardly realized in such a uniform
Bose-Hubbard system. In this paper, we present three examples to demonstrate
how to realize the perfect coherent shift in both boson and fermion systems.
We present two types of local BP for Bose system and one type for Fermi
system. It is shown exactly that the perfect coherent shift can be achieved
in such models. We find that for a Bose on-site BP, the perfect coherent
shift condition depends on the nearest-neighbor (NN) interaction strength
and the momentum of the incident single particle wavepacket, while for the
other two types of BPs, it is independent of the initial state in the
proposed systems. Our results indicate that the coherent shift phenomenon
has a great potential for future applications.

The article is organized as follows: In Sec. II, we introduce an extended
Bose-Hubbard and the formation of the two particle bound state. In Sec. III,
we investigate the coherent shift process for an on-site BP in the case of
weak NN interaction. Sec. IV is devoted to the same discussion for the NN
BP. In Sec. V, we introduce the on-site BP in the Fermi system and
demonstrate how to perform a perfect coherent shift. Sec. VI is the
conclusion and a short discussion.

\section{Extended Bose-Hubbard model}

We begin with the Bose on-site BP, considering the scattering process
between it and a single boson in an extended Bose-Hubbard model. The
Hamiltonian reads

\begin{eqnarray}
H^{\text{B}} &=&-\kappa \sum_{i}\left( a_{i}^{\dag }a_{i+1}+\text{H.c.}%
\right) +\frac{U}{2}\sum_{i}n_{i}\left( n_{i}-1\right)
\label{extend Hubbard} \\
&&+V\sum_{i}n_{i}n_{i+1},  \notag
\end{eqnarray}%
where $a_{i}^{\dag }$ $\left( a_{i}\right) $ is the particles creation
(annihilation) operator and $n_{i}=a_{i}^{\dag }a_{i}$ is the number
operator at the $i$th lattice site. The tunneling strength and the on-site
interaction between bosons are denoted by $\kappa $ and $U$. Hamiltonian Eq.(%
\ref{extend Hubbard}) has an additional NN interaction $V$ in comparison to
the original Hamiltonian Eq. (1) in previous paper \cite{JLBP}.

First of all, a state in the two-particle Hilbert space, as shown in Ref.
\cite{JLBP}, can be written as
\begin{equation}
\left\vert \psi _{k}\right\rangle =\sum_{k,r}f^{k}(r)\left\vert \phi
_{r}^{k}\right\rangle ,  \label{Psi_k}
\end{equation}%
with\
\begin{eqnarray}
\left\vert \phi _{0}^{k}\right\rangle &=&\frac{1}{\sqrt{2N}}%
\sum_{j}e^{ikj}\left( a_{j}^{\dag }\right) ^{2}\left\vert \text{vac}%
\right\rangle ,  \label{phi_k} \\
\left\vert \phi _{r}^{k}\right\rangle &=&\frac{1}{\sqrt{N}}%
e^{ikr/2}\sum_{j}e^{ikj}a_{j}^{\dag }a_{j+r}^{\dag }\left\vert \text{vac}%
\right\rangle ,  \notag
\end{eqnarray}%
here $\left\vert \text{vac}\right\rangle $\ is the vacuum state for the
boson operator $a_{i}$. $k=2\pi n/N$, $n\in \lbrack 1,N]$ denotes the
momentum with $N=2N_{0}+1$, and $r\in \lbrack 1,N_{0}]$ is the distance
between the two particles. The matrix representation for the Hamiltonian
operator $H^{\text{B}}$ in the basis $\{\left\vert \phi
_{j}^{k}\right\rangle \}$ is

\begin{equation}
H^{k}=\left(
\begin{array}{ccccc}
U & \sqrt{2}T^{k} &  &  &  \\
\sqrt{2}T^{k} & V & T^{k} &  &  \\
& T^{k} &  & \ddots &  \\
&  & \ddots &  & T^{k} \\
&  &  & T^{k} & T_{N_{0}}^{k}%
\end{array}%
\right) ,  \label{Hk}
\end{equation}%
where $T^{k}=-2\kappa \cos \left( k/2\right) $, and $T_{N_{0}}^{k}=\left(
-1\right) ^{n}T^{k}$. Note that for an arbitrary $k$, the eigenvalues of (%
\ref{Hk}) are equivalent to that of the single-particle $N_{0}+1$-site
tight-binding chain system with one abnormal NN hopping amplitude $\sqrt{2}%
T^{k}$ between $0$th ($\left\vert \phi _{0}^{k}\right\rangle $)\ and $1$st
site ($\left\vert \phi _{1}^{k}\right\rangle $), and the on-site potentials $%
U$, $V$ and $T_{N_{0}}^{k}$\ at $0$th, $1$st and $N_{0}$th sites
respectively. It follows that in each $k$-invariant subspace, there are
three types of bound states arising from the on-site potentials under the
following conditions. In the case with $\left\vert U-V\right\vert \gg
\left\vert \kappa \right\vert $,\ the particle can be localized at either $0$%
th\ or $1$st\ site, corresponding to (i) the on-site BP state, or (ii) the
NN BP state. Interestingly, in the case of $U=V$, and $\left\vert
U\right\vert $, $\left\vert V\right\vert \gg \left\vert \kappa \right\vert $%
, the particle can be in the bonding state (or anti-bonding state) between $%
0 $th\ and $1$st\ sites, which was discussed in Ref. \cite{UVBS}. In
previous works \cite{JLBP, MVExtendB}, the bound states of (i) and (ii) were
well investigated. In Ref. \cite{JLBP}, it was shown that the coherent shift
of an on-site BP occurs when it meets a single particle. In the following we
focus on the scattering process between a BP and a single particle in the
presence of the NN interaction. We will show exactly that the NN interaction
can lead to the\ perfect coherent shift.

\begin{figure}[tbp]
\includegraphics[ bb=32 282 425 782, width=6.0 cm, clip]{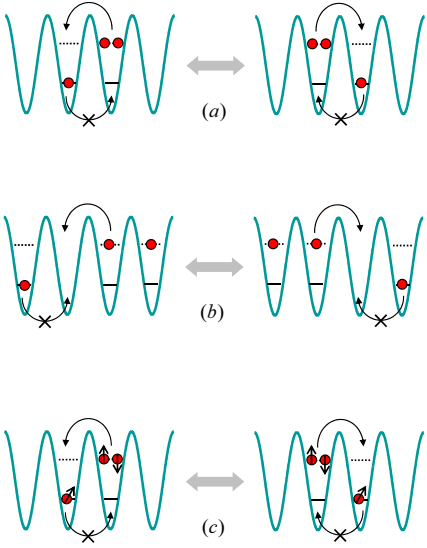}
\caption{(Color online) Schematic illustrations of the swapping processes
leading to the perfect coherent shift in both Bose and Fermi systems. (a)
On-site BP in an extended Bose-Hubbard model. (b) NN BP in an extended
Bose-Hubbard model. (c) On-site singlet BP in a simple Fermi-Hubbard model.}
\label{swap}
\end{figure}


\section{On-site bound pair in Bose system}

Now we start with the on-site BP state. The formation of bound pair state in
the Bose-Hubbard system was studied in Ref. \cite{Winkler, Valiente, JLBP,
MVExtendB}. Here in the extended Bose-Hubbard model, the on-site BP
corresponds to the BP bounded by $U$ in the case of large $U$\ but weak $V$,
with $\left\vert U\right\vert \gg \left\vert \kappa \right\vert \sim
\left\vert V\right\vert $. In this case, the solution of $f^{k}(r)$\ has the
form%
\begin{equation}
f^{k}(r)\simeq \left\{
\begin{array}{c}
1\text{,\ }(r=0), \\
0\text{,\ }(r\neq 0),%
\end{array}%
\right. ,
\end{equation}%
with eigenenergy%
\begin{equation}
\varepsilon _{k}\simeq U+\frac{4\kappa ^{2}}{U}\left( \cos k+1\right) .
\end{equation}%
One can see that the on-site BP state acts as a composite particle with
effective hopping strength being $2\kappa ^{2}/U$. According to Ref. \cite%
{JLBP}, the effective Hamiltonian for a single on-site BP and a single
particle\ in the extended Bose-Hubbard system can be obtained in the form%
\begin{eqnarray}
H_{\text{eff}}^{\text{U}} &=&-\kappa \sum_{i}\left( \tilde{a}_{i}^{\dag }%
\tilde{a}_{i+1}+2\tilde{b}_{i+1}^{\dag }\tilde{b}_{i}\tilde{a}_{i}^{\dag }%
\tilde{a}_{i+1}+\text{H.c.}\right)  \label{Heff_b1} \\
&&+\frac{2\kappa ^{2}}{U}\sum_{i}\left( \tilde{b}_{i}^{\dag }\tilde{b}_{i+1}+%
\text{H.c.}\right)  \notag \\
&&+\left( 2V-\frac{7\kappa ^{2}}{2U}\right) \sum_{i}\tilde{b}_{i}^{\dag }%
\tilde{b}_{i}\left( \tilde{a}_{i-1}^{\dag }\tilde{a}_{i-1}+\tilde{a}%
_{i+1}^{\dag }\tilde{a}_{i+1}\right)  \notag \\
&&+\left( U+\frac{4\kappa ^{2}}{U}\right) \sum_{i}\tilde{b}_{i}^{\dagger }%
\tilde{b}_{i},  \notag
\end{eqnarray}%
where $\tilde{a}_{i}$\ and $\tilde{b}_{i}$\ denote the hardcore bosons
satisfying the following commutation relations $[\tilde{a}_{j},\tilde{a}%
_{i}^{\dag }]=$ $[\tilde{b}_{j},\tilde{a}_{i}^{\dag }]=$ $[\tilde{b}_{j},%
\tilde{b}_{i}^{\dag }]=0,$ $(i\neq j)$; $\{\tilde{a}_{i},\tilde{a}_{i}^{\dag
}\}=\{\tilde{b}_{i},\tilde{b}_{i}^{\dag }\}=1$; $\{\tilde{b}_{i},\tilde{a}%
_{i}^{\dag }\}=\{\tilde{b}_{i},\tilde{a}_{i}\}=0$. The former two terms of $%
H_{\text{eff}}^{\text{U}}$ depict the hoppings while the last two terms
depict the interaction bewtween the two kinds of particles. Considering the
scattering problem without loss of generality, we set the particle $\tilde{b}%
^{\dag }\left\vert \text{vac}\right\rangle $ at the $0$th site and incident
the particle $\tilde{a}^{\dag }\left\vert \text{vac}\right\rangle $ from $%
-\infty $ at the beginning. For the scattering process between them within
short duration, the particle $\tilde{b}^{\dag }\left\vert \text{vac}%
\right\rangle $ is static compared to the single particle $\tilde{a}^{\dag
}\left\vert \text{vac}\right\rangle $. Then the whole scattering process
dominantly governed by

\begin{eqnarray}
\mathcal{H}_{\text{eff}}^{\text{U}} &=&-\kappa \sum_{i=-\infty }\tilde{a}%
_{i}^{\dag }\tilde{a}_{i+1}-2\kappa \tilde{b}_{-1}^{\dag }\tilde{b}_{0}%
\tilde{a}_{0}^{\dag }\tilde{a}_{-1}+\text{H.c.}  \label{H_scatt} \\
&&+2V(\tilde{a}_{-1}^{\dag }\tilde{a}_{-1}\tilde{b}_{0}^{\dag }\tilde{b}_{0}+%
\tilde{b}_{-1}^{\dag }\tilde{b}_{-1}\tilde{a}_{0}^{\dag }\tilde{a}_{0})
\notag \\
&&+U\tilde{b}_{0}^{\dagger }\tilde{b}_{0}+U\tilde{b}_{-1}^{\dagger }\tilde{b}%
_{-1},  \notag
\end{eqnarray}%
at the condition $\left\vert U\right\vert \gg \left\vert \kappa \right\vert
\sim \left\vert V\right\vert $. Here we neglected the terms with $\kappa
^{2}/U$ and $\kappa ^{2}/V$. The swapping process between a single particle
and a on-site BP is the key of the coherent shift, which is schematically
illustrated in Fig. \ref{swap}(a). The scattering process is represented in
the form

\begin{equation}
\tilde{a}_{-\infty }^{\dag }\tilde{b}_{0}^{\dag }\left\vert \text{vac}%
\right\rangle \rightarrow r\tilde{a}_{-\infty }^{\dag }\tilde{b}_{0}^{\dag
}\left\vert \text{vac}\right\rangle +t\tilde{a}_{\infty }^{\dag }\tilde{b}%
_{-1}^{\dag }\left\vert \text{vac}\right\rangle ,  \label{scattering}
\end{equation}%
where $r$ and $t$ are the reflection and transmission (coherent shift)
amplitudes, respectively.

\begin{figure}[tbp]
\includegraphics[ bb=32 620 418 763, width=6.0 cm, clip]{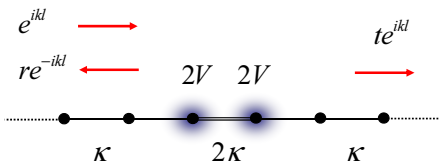}
\caption{(Color online) Schematic illustration of the equivalent effective
Hamiltonian governs the evolution of a single particle and an on-site BP. It
is a chain with an embedded impurity, which consists of two neighboring
sites with identical on-site potentials $2V$ and the tunneling strength $2%
\protect\kappa $\ between them. The resonant transmission occurs at $V=V_{R}$%
, and is equivalent to the perfect coherent shift.}
\label{single}
\end{figure}

In order to investigate the above Hamiltonian Eq. (\ref{H_scatt}), we define
a new set of basis $\{\left\vert l\right\rangle _{u}\}$ as%
\begin{equation}
\left\vert l\right\rangle _{u}\equiv \left\{
\begin{array}{r}
\tilde{a}_{l}^{\dag }\tilde{b}_{0}^{\dag }\left\vert \text{vac}\right\rangle
=\tilde{a}_{l}^{\dag }\tilde{a}_{0}^{\dag 2}/\sqrt{2}\left\vert \text{vac}%
\right\rangle \text{, }(l<0) \\
\tilde{b}_{-1}^{\dag }\tilde{a}_{l}^{\dag }\left\vert \text{vac}%
\right\rangle =\tilde{a}_{-1}^{\dag 2}\tilde{a}_{l}^{\dag }/\sqrt{2}%
\left\vert \text{vac}\right\rangle \text{,\ }(l\geqslant 0)%
\end{array}%
\right. ,  \label{BP_basis}
\end{equation}%
and then reduce the two-body problem to a single particle problem. Actually,
acting the Hamiltonian of Eq. (\ref{H_scatt})\ or the original Hamiltonian
Eq. (\ref{extend Hubbard}) on the basis (\ref{BP_basis}), we obtain the
equivalent single particle Hamiltonian

\begin{eqnarray}
H_{\text{sp}} &=&-\kappa \left( \sum_{l=-\infty }^{-2}+\sum_{l=0}^{+\infty
}\right) \left\vert l\right\rangle _{u}\left\langle l+1\right\vert -2\kappa
\left\vert -1\right\rangle _{u}\left\langle 0\right\vert +\text{H.c.}  \notag
\\
&&+2V\left( \left\vert -1\right\rangle _{u}\left\langle -1\right\vert
+\left\vert 0\right\rangle _{u}\left\langle 0\right\vert \right) \text{.}
\label{single_P}
\end{eqnarray}%
The physics of the equivalent Hamiltonian is obvious, which describes a
particle in the chain with an embedded impurity. The impurity consists of
two neighboring sites with identical on-site potentials $2V$ and the
tunneling strength $2\kappa $\ between them. Fig. \ref{single} is a
schematic illustration for the equivalent Hamiltonian.\ Then the scattering
process of Eq. (\ref{scattering}) can be rewritten as%
\begin{equation}
a_{-\infty }^{\dag }\left\vert \text{vac}\right\rangle \rightarrow
ra_{-\infty }^{\dag }\left\vert \text{vac}\right\rangle +ta_{\infty }^{\dag
}\left\vert \text{vac}\right\rangle .  \label{scattering 1}
\end{equation}%
The transmission coefficient $\left\vert t\right\vert ^{2}$ corresponds to
the success probability of the coherent shift. For an incident plane wave of
momentum $k$, its expression as a function of $k$ and $V$\ can be derived by
using the Green's function method \cite{Datta,YangAs} or the Bethe-ansatz
technique \cite{JLTrans}.

The retarded Green's function of the system of Eq. (\ref{single_P}) for an
input energy $E=-2\kappa \cos k$ is

\begin{equation}
G^{R}=\frac{1}{E-H_{c}-\sum^{R}},  \label{G_R}
\end{equation}%
where $H_{c}$ is the Hamiltonian of the central system%
\begin{equation}
H_{c}=\left(
\begin{array}{cc}
2V & -2\kappa \\
-2\kappa & 2V%
\end{array}%
\right) ,
\end{equation}%
and $\sum^{R}$ denotes the contribution of the half-infinite leads which can
be viewed as an effective Hamiltonian arising from the interaction of the
central system with the leads%
\begin{equation}
\sum\nolimits^{R}=\left(
\begin{array}{cc}
-\kappa e^{ik} & 0 \\
0 & -\kappa e^{ik}%
\end{array}%
\right) .
\end{equation}%
The transmission probability is given by

\begin{equation}
\left\vert t\right\vert ^{2}=T_{12}=\text{Tr}\left[ \Gamma _{1}G^{R}\Gamma
_{2}G^{A}\right] .
\end{equation}%
where the advanced Green's function $G^{A}$ and $\Gamma _{1,2}$ matrices are
\begin{equation}
\begin{array}{c}
G^{A}=G^{R\dag }\text{,} \\
\Gamma _{1}=\left(
\begin{array}{cc}
2\kappa \sin k & 0 \\
0 & 0%
\end{array}%
\right) \text{,} \\
\Gamma _{2}=\left(
\begin{array}{cc}
0 & 0 \\
0 & 2\kappa \sin k%
\end{array}%
\right) .%
\end{array}%
\end{equation}%
Then we can obtain the transmission probability as,%
\begin{equation}
T_{12}=\frac{16\sin ^{2}k}{\prod\limits_{l=-1,1}\left[ 4\left( V/\kappa
+l\right) ^{2}+4\left( V/\kappa +l\right) \cos k+1\right] }.
\end{equation}%
Noting that the transmission coefficient is $k$ dependent, we calculate the
resonant transmission or perfect coherent shift. For $T_{12}=1$, we have the
NN coupling constant $V=V_{R}$, where

\begin{equation}
V_{R}=\frac{\kappa }{2}\left( -\cos k\pm \sqrt{\cos ^{2}k+3}\right) .
\label{VR}
\end{equation}%
It is worth noting that Eq. (\ref{VR}) is available for any $k$,\ since $%
V_{R}$\ is always in the order of $\kappa $.\ In practice, this process can
be implemented via a broad wavepacket. For the case of $k=\pi /2$, which
corresponds to the stablest and fastest wave packet \cite{KimImpurity},\ we
can generate a unitary swap between an on-site BP and a single particle
under the condition $V_{R}=\pm \sqrt{3}\kappa /2$. In this condition, the $k$
dependent transmission coefficient\ $T_{12}\left( k\right) =4\sin
^{2}k/\left( 4-\cos ^{2}k\right) $. Then we get the conclusion that the
perfect coherent shift can be achieved in the large $U$ Bose-Hubbard model
with weak NN interaction.

\section{NN bound pair in Bose system}

Now we consider another kind of bound pair in the extended Bose-Hubbard
model (\ref{extend Hubbard}). This bound pair is bounded by the NN
interaction $V$, rather than the on-site interaction $U$. In large $V$
limit, $\left\vert V\right\vert $,$\ \left\vert V-U\right\vert \gg
\left\vert \kappa \right\vert $, a pair of hardcore bosons can be bounded by
the NN interaction $V$. This was pointed out by M. Valiente \textit{et al}.
in their work \cite{MVExtendB}. In such condition, the dynamics of a single
boson and an NN BP can be depicted by the following effective Hamiltonian%
\begin{eqnarray}
H_{\text{eff}}^{\text{V}} &=&-\kappa \sum_{i}\left( a_{i}^{\dag
}a_{i+1}+B_{i}^{\dag }B_{i+2}a_{i+3}^{\dag }a_{i}+\text{H.c.}\right)
\label{Heff_V} \\
&&+\left( \frac{\kappa ^{2}}{V}+\frac{2\kappa ^{2}}{V-U}\right)
\sum_{i}\left( B_{i}^{\dag }B_{i+1}+\text{H.c.}\right)  \notag \\
&&-\frac{2\kappa ^{2}}{V}\sum_{i}\left( B_{i}^{\dag }B_{i}a_{i-2}^{\dag
}a_{i-2}+B_{i}^{\dag }B_{i}a_{i+3}^{\dag }a_{i+3}\right)  \notag \\
&&+\left( V+\frac{2\kappa ^{2}}{V}+\frac{4\kappa ^{2}}{V-U}\right)
\sum_{i}B_{i}^{\dagger }B_{i},  \notag
\end{eqnarray}%
where the NN pair operator is defined as $B_{i}^{\dag }=a_{i}^{\dag
}a_{i+1}^{\dag }$. For the scattering process between the NN BP and the
single particle within short duration, it is dominantly governed by the
first term, which includes the hopping of a single particle and the swapping
between them. The swapping process is schematically illustrated in Fig. \ref%
{swap}(b). We consider the scattering problem between a single particle and
an NN BP. Initially, an NN BP $B_{2}^{\dag }\left\vert \text{vac}%
\right\rangle =a_{2}^{\dag }a_{3}^{\dag }\left\vert \text{vac}\right\rangle $
is located at the dimer of sites $2$ and $3$, while a single particle $%
a^{\dag }\left\vert \text{vac}\right\rangle $ is located at the left.
Similarly, we can define a set of basis $\{\left\vert l\right\rangle _{v}\}$
as%
\begin{equation}
\left\vert l\right\rangle _{v}\equiv \left\{
\begin{array}{r}
a_{l}^{\dag }B_{2}^{\dag }\left\vert \text{vac}\right\rangle =a_{l}^{\dag
}a_{2}^{\dag }a_{3}^{\dag }\left\vert \text{vac}\right\rangle \text{, }%
(l\leqslant 0) \\
B_{0}^{\dag }a_{l}^{\dag }\left\vert \text{vac}\right\rangle =a_{0}^{\dag
}a_{1}^{\dag }a_{l+2}^{\dag }\left\vert \text{vac}\right\rangle \text{,\ }%
(l>0)%
\end{array}%
\right. .  \label{V_BP_basis}
\end{equation}%
Acting the Hamiltonian Eq.(\ref{Heff_V}) on the basis Eq.(\ref{V_BP_basis}),
after neglecting the high order terms $\kappa ^{2}/V$ and $\kappa
^{2}/\left( V-U\right) $, we obtain a unifirm tight-binding chain.
Obviously, such a system can realize the perfect coherent shift for any
incident single particle wave with the shift distance being \textit{two
lattice spacings}. Comparing to the previous on-site BP, the perfect
coherent shift of the NN BP is $k$ independent.

\section{On-site bound pair in Fermi system}

As pointed out in Ref. \cite{Comment}, the nonzero reflection of a single
incident particle in the coherent shift process \cite{JLBP} is attributed to
the swapping strength\ being $2\kappa $ rather than $\kappa $. Essentially,
this arises from the identity of two particles of the bound pair.
Consequently, for a system with the bound pair consists of two particles
with opposite spins, the unexpected reflection should be avoidable.

Now we turn to the Fermi system. A one-dimensional Fermi-Hubbard Hamiltonian
reads
\begin{equation}
H^{\text{F}}=-\kappa \sum_{i,\sigma }\left( c_{i,\sigma }^{\dagger
}c_{i+1,\sigma }+\text{H.c.}\right) +U\sum_{i}n_{i\uparrow }n_{i\downarrow },
\label{F_Hub}
\end{equation}%
where $c_{i,\sigma }^{\dagger }$ is the creation operator of the fermion at
the site $i$ with spin $\sigma =\uparrow ,\downarrow $ and $U$ is the
on-site interaction. Similarly, there also exists bound pair state in such a
system. Actually, a state in the two-particle Hilbert space with spin zero
can be written as the form of Eq. (\ref{Psi_k}), where we redefine the
corresponding basis as

\begin{eqnarray}
\left\vert \phi _{0}^{k}\right\rangle ^{\text{F}} &=&\frac{1}{\sqrt{N}}%
\sum_{j}e^{ikj}c_{j,\uparrow }^{\dagger }c_{j,\downarrow }^{\dagger
}\left\vert \text{vac}\right\rangle ,  \notag \\
\left\vert \phi _{r}^{k}\right\rangle ^{\text{F}} &=&\frac{1}{\sqrt{2N}}%
e^{ikr/2}\sum_{j}e^{ikj} \\
&&\times \left( c_{j,\uparrow }^{\dagger }c_{j+r,\downarrow }^{\dagger
}-c_{j,\downarrow }^{\dagger }c_{j+r,\uparrow }^{\dagger }\right) \left\vert
\text{vac}\right\rangle .  \notag
\end{eqnarray}%
Then all the analysis for the formation of a Bose on-site BP can be applied
completely on that of a Fermi on-site BP. Besides, in the large $U$ limit,
the effective Hamiltonian describing the dynamics of a single particle and a
Fermi on-site BP has the form
\begin{eqnarray}
H_{\text{eff}}^{\text{F}} &=&-\kappa \sum_{i,\sigma }\left( \tilde{c}%
_{i,\sigma }^{\dagger }\tilde{c}_{i+1,\sigma }+\tilde{c}_{i,\sigma
}^{\dagger }\tilde{c}_{i+1,\sigma }d_{i+1}^{\dag }d_{i}+\text{H.c.}\right)
\notag \\
&&+\frac{2\kappa ^{2}}{U}\sum_{i}\left( d_{i}^{\dagger }d_{i+1}+\text{H.c.}%
\right)  \label{H_eff_F} \\
&&-\frac{2\kappa ^{2}}{U}\sum_{i,\sigma }\tilde{c}_{i,\sigma }^{\dagger }%
\tilde{c}_{i,\sigma }\left( d_{i-1}^{\dagger }d_{i-1}+d_{i+1}^{\dagger
}d_{i+1}\right)  \notag \\
&&+\left( U+\frac{4\kappa ^{2}}{U}\right) \sum_{i}d_{i}^{\dagger }d_{i},
\notag
\end{eqnarray}%
where $\tilde{c}_{i,\sigma }=c_{i,\sigma }\left( 1-n_{i,-\sigma }\right) $
is the projected fermion creation operator, $d_{i}=c_{i,\downarrow
}c_{i,\uparrow }$\ is the on-site BP operator. The projector $\left(
1-n_{i,-\sigma }\right) $ allows to create an electron with spin $\sigma $\
at the site $i$ only if there is no other electron on that site. For the
scattering process between $\tilde{c}_{i,\sigma }$ and $d_{i}$ within short
duration, it is dominantly governed by the first term, which includes the
hopping of the single particle and the swapping between them. The swapping
process is schematically illustrated in Fig. \ref{swap}(c). We note that the
swapping operation $\tilde{c}_{i,\sigma }^{\dagger }\tilde{c}_{i+1,\sigma
}d_{i+1}^{\dag }d_{i}$ has the same coupling strength with the term $\tilde{c%
}_{i,\sigma }^{\dagger }\tilde{c}_{i+1,\sigma }$. This allows the perfect
coherent shift. In addition, such a process is independent of the momentum
of the incident particle and the spin polarization, since the bound pair is
singlet.

\section{Conclusion}

In summary, we present three kinds of bound pairs and the corresponding
optimal systems, which can avoid unexpected reflection in the coherent shift
process. It is shown exactly that the perfect coherent shift can be achieved
in the simply engineered systems. For a Bose on-site BP, the perfect
coherent shift requires the resonant condition, which depends on the NN
interaction strength and the momentum of the incident single particle
wavepacket. For a Bose NN BP and a Fermi on-site BP, the perfect coherent
shifts occur for an arbitrary initial state in the simple chain systems. We
believe that our findings have a great potential for future applications.

We acknowledge the support of the CNSF (Grant Nos. 10874091 and
2006CB921205).


\begin{thebibliography}{99}
\bibitem{Winkler} K. Winkler, G. Thalhammer, F. Lang, R. Grimm, J. H.
Denschlag, A. J. Daley, A. Kantian, H. P. B\"{u}chler and P. Zoller, Nature
(London) \textbf{441}, 853 (2006).

\bibitem{Mahajan} S. M. Mahajan and A. Thyagaraja, J. Phys. A \textbf{39},
L667 (2006).

\bibitem{Petrosyan} D. Petrosyan, B. Schmidt, J. R. Anglin, and M.
Fleischhauer, Phys. Rev. A \textbf{76}, 033606 (2007).

\bibitem{Creffield} C. E. Creffield, Phys. Rev. A \textbf{75}, 031607(R)
(2007).

\bibitem{Kuklov} A. Kuklov and H. Moritz, Phys. Rev. A \textbf{75}, 013616
(2007).

\bibitem{Folling} D. Petrosyan, B. Schmidt, J. R. Anglin, and M.
Fleischhauer, Phys. Rev. A \textbf{76}, 033606 (2007).

\bibitem{Zollner} S. Z\"{o}llner, H.-D. Meyer, and P. Schmelcher, Phys. Rev.
Lett. \textbf{100}, 040401 (2008).

\bibitem{ChenS} L. Wang, Y. Hao, and S. Chen, Eur. Phys. J. D \textbf{48},
229 (2008).

\bibitem{Valiente} M. Valiente and D. Petrosyan, J. Phys. B \textbf{41},
161002 (2008).

\bibitem{JLBP} L. Jin, B. Chen and Z. Song, Phys. Rev. A \textbf{79}, 032108
(2009).

\bibitem{MVExtendB} M. Valiente and D. Petrosyan, J. Phys. B \textbf{42},
121001 (2009).

\bibitem{Comment} M. Valiente, D. Petrosyan, and A. Saenz, Comment on
\textquotedblleft Coherent shift of localized bound pairs in the
Bose-Hubbard model\textquotedblright .

\bibitem{UVBS} L. Jin, Z. Song, arXiv:1004.4770v1.

\bibitem{Datta} S. Datta, \textit{Electronic Transport in Mesoscopic Systems}
(Cambridge University Press, Cambridge, 1995).

\bibitem{YangAs} S. Yang, Z. Song and C. P. Sun, arXiv:0912.0324v1.

\bibitem{JLTrans} L. Jin and Z. Song, Phys. Rev. A \textbf{81}, 022107
(2010).

\bibitem{KimImpurity} W. Kim, L. Covaci, and F. Marsiglio, Phys. Rev. B
\textbf{74}, 205120 (2006).
\end{thebibliography}
\end{document}